\documentclass[aps,prl,reprint,superscriptaddress,nofootinbib]{revtex4-2}
\usepackage{amsmath,amssymb,bm}
\usepackage{hyperref}
\usepackage{microtype}

\newcommand{\PBZ}{P_{\rm BZ}}
\newcommand{\OH}{\Omega_H}

\begin{document}
\title{Irreducible-Mass Growth and Extractable Energy in a Self-Gravitating Blandford--Znajek Engine}
\author{Remo Ruffini}
\email{ruffini@icra.it}
\affiliation{International Center for Relativistic Astrophysics Network (ICRANet)\\
Piazza della Repubblica 10, 65122 Pescara, Italy}
\author{Giorgio Sonnino}
\email{giorgio.sonnino@ulb.be}
\affiliation{Université Libre de Bruxelles (ULB)\&\\
International SOLVAY Institutes for Physics and Chemistry\\
Campus de la Plaine, P.C. 224, Build. NO, Bvd. du Triomphe, Brussels 1050, Belgium
}
%\date{\today}

\begin{abstract}
Wang recently constructed a self-gravitating split-monopole Blandford-Znajek engine and obtained its secular horizon mass and angular-momentum loss rates. We determine the previously unknown consequences of this evolution: the growth of the irreducible mass, the division of rotational work between outgoing electromagnetic power and horizon dissipation, and the lifetime-integrated energy removed from the horizon. For fixed magnetic flux and impedance matching, one half of the instantaneous rotational work is carried outward, and one half irreversibly increases the horizon area. Integrating Wang's spin-down trajectory, we find in the weak-field limit the exact results $A_f/A_0=\sqrt e$ and $E_{\rm extr}^{H}/(M_{H,0}c^2)=1-e^{1/4}/\sqrt2\simeq0.09206$. The leading maximum-power spin-flux allocation instead gives $E_{\rm extr}^{H}/(M_{H,0}c^2)\simeq0.04668$. These results provide the first irreducible-mass and extractable-energy accounting of Wang's engine. Their extension from quasilocal horizon quantities to energy received at infinity requires a separate flux analysis.
\end{abstract}
\maketitle

We use geometrized units $G=c=1$, restoring $c$ where indicated.

\noindent \emph{Introduction}

The fundamental Black Holes physics and astrophysics are rooted in 3 papers, in which the mass-energy of the Black Holes is related to 1) its irreducible mass $M_{\rm irr}$ \cite{Christodoulou}, 2) to the surface area of the horizon \cite{ruffini1} 3) monotonically increasing both in the case of an isolated Black Hole or a merging binary system \cite{Hawking1971}.

\noindent Ruffini and Wilson first demonstrated that frame dragging twists the magnetic field of plasma accreting onto a Kerr black hole, producing a torque capable of extracting rotational energy and inducing magnetospheric charge separation \cite{ruffini}. The Blandford-Znajek (BZ) mechanism realizes this extraction through a horizon-threading, force-free magnetosphere \cite{BZ1977,MT1982}. In the standard treatment, the electromagnetic field is a test field on Kerr, so its stress-energy does not participate in the geometry.

\noindent Wang introduces a self-gravitating alternative in which the electromagnetic stress-energy is included consistently in the spacetime \cite{Wang2026}. This construction provides a useful framework for examining the Blandford-Znajek process. Our purpose is to use Wang's approach to determine the previously unexplored evolution of the irreducible mass, clarify the division of rotational work between electromagnetic extraction and irreversible horizon growth, and derive the corresponding lifetime-integrated energy removed from the horizon. 

\noindent In the Wang solution, the exact static seed joins magnetic Reissner-Nordstr\"om hemispheres with parameters $+P$ and $P$. Their metrics agree because they depend on $P^2$, while an equatorial current sheet supports the reversal of the electromagnetic field. Each hemisphere carries flux $\Phi_{N,S}=\pm2\pi P$, but the total magnetic charge vanishes. The seed metric for a black hole of mass $M$ is
\begin{equation}\label{I1}
f(r)=1-\frac{2M}{r}+\frac{P^2}{r^2}
\end{equation} 
and is invariant under $P\rightarrow-P$. Rotation is introduced perturbatively through the spin of the black hole $a=J_H/M_H$ and the angular velocity of the black-hole horizon $\Omega_H=a/(r_+^2+a^2)$, where $r_+$ is the radial coordinate of the outer event horizon: $r_+=M_H+\sqrt{M_H^2-a^2-P^2}$. Here $M_H(t)$ and $J_H(t)$ are the quasilocal mass and the quasilocal angular momentum associated with the black-hole horizon, respectively. Denoting $\Omega_F$ as the angular velocity of the magnetic-field lines, a stationary regular horizon enforces $\Omega_F=\Omega_H$,  and no extraction, whereas matching the quasistationary horizon to an outgoing load selects $\Omega_F=\Omega_H/2$. The horizon then evolves according to
\begin{equation}
\dot M_H=-P_{\rm BZ}, \qquad \dot J_H=-\frac{P_{\rm BZ}}{\Omega_F} =-\frac{2P_{\rm BZ}}{\Omega_H}\label{eq:bzrates}
\end{equation} 
where $P_{\rm BZ}=dE_{\rm extr}^{H}/dt>0$ is the electromagnetic power removed from the horizon. The static seed and stationary-horizon obstruction are exact; the extracting engine is a slow-rotation, quasistationary solution of the coupled Einstein--Maxwell--force-free system rather than a global exact solution at arbitrary spin. Wang determined the evolution rates in Eq.~\eqref{eq:bzrates}, but not their thermodynamic consequences. Here we derive the associated irreducible-mass growth, identify the irreversible partition of rotational work, and integrate the spin-down trajectory to obtain the total energy removed from the horizon. The distinction between this quasilocal energy and the energy received at infinity is essential because the hemispheric flux is supported by a current sheet rather than by a conserved magnetic charge.

\noindent\emph{Generalized irreducible-mass theorem}

\noindent Let a nonextremal, quasistationary, axisymmetric horizon be described by its quasilocal mass $M_H(t)$, the angular momentum $J_H(t)$, the area $A(t)$, the surface gravity $\kappa(t)$, and the hemispheric flux $\Phi(t)$. We assume that neighboring horizon states admit an isolated-horizon Hamiltonian satisfying
\begin{equation}
 \dot M_H=\frac{\kappa}{8\pi}\dot A +\OH\dot J_H+\frac{\phi_H}{2\pi}\dot\Phi . \label{eq:firstlaw}
\end{equation}
with $\phi_H$ denoting the horizon magnetic potential conjugate to $P$. For a genuine electromagnetic charge, this structure follows from isolated-horizon mechanics \cite{Ashtekar2001}. For Wang's sheet-supported hemispheric flux, however, Eq.~\eqref{eq:firstlaw} is a conditional quasilocal assumption: its derivation requires the horizon Hamiltonian and the equatorial current-sheet boundary term. It must not be identified with a variation of the Arnowitt-Deser-Misner (ADM) or Bondi mass.
Defining the irreducible mass $M_{\rm irr}$ as
\begin{equation}
 M_{\rm irr}\equiv\sqrt{\frac{A}{16\pi}}, \qquad {\dot A}=32\pi M_{\rm irr}{\dot M}_{\rm irr}\label{eq:mirrdef}
\end{equation}
Eq.~\eqref{eq:firstlaw} gives
\begin{equation}
\dot M_{\rm irr}= \frac{\dot M_H-\OH\dot J_H-(\phi_H/2\pi)\dot\Phi}  {4\kappa M_{\rm irr}}\equiv\frac{\mathcal D_H}{4\kappa M_{\rm irr}}
 \label{eq:mirrtheorem}
\end{equation}
This is the generalized irreducible-mass balance. So, the black-hole mass balance separates into rotational work, magnetic work,
and irreversible horizon dissipation:
\begin{equation}
 \dot M_H=\underbrace{\Omega_H\dot J_H}_{\rm rotational\ work}+\underbrace{\frac{\phi_H}{2\pi}\dot\Phi}_{\rm magnetic\ work}+\underbrace{\mathcal D_H}_{\rm horizon\ dissipation}\label{eq:mass-balance}
\end{equation}
Here $\mathcal D_H$ is not an independent energy source; it is the remainder of the total energy flux after subtracting the rotational and magnetic work terms:
\begin{equation}
 \mathcal D_H \equiv  \dot M_H-\Omega_H\dot J_H-\frac{\phi_H}{2\pi}\dot\Phi=\frac{\kappa}{8\pi}\dot A \label{eq:horizon-dissipation}
\end{equation}
It is therefore the irreversible contribution responsible for the growth of the horizon area. From Eq.~(\ref{eq:mirrdef}) we get
\begin{equation}
 \dot M_{\rm irr}=\frac{\mathcal D_H}{4\kappa M_{\rm irr}}\label{eq:mirr-growth}
\end{equation}
The standard physical-process first law relates variations of Hamiltonian charges to fluxes through the horizon. Replacing those charges by $M_H$ and $J_H$ is justified only after specifying the corresponding isolated-horizon Hamiltonian. Subject to the assumption in Eq.~\eqref{eq:firstlaw}, the integrated dissipative combination is the canonical energy absorbed by the horizon. At linear order, it is determined by the stress-energy flux,
\begin{equation}\label{D}
 \Delta E_H^{\rm diss}\!=\!\Delta M_H\!-\!\Omega_H\Delta J_H\!-\!\phi_H\Delta P\!=\!\frac{\kappa}{8\pi}\Delta A\!=\!\int_{\mathcal H}T_{\mu\nu}\chi^\mu d\Sigma^\nu 
\end{equation}
where $\Sigma^\nu$ is the directed three-dimensional surface element on the horizon. In Eq.~(\ref{D}), $T_{\mu\nu}$ is the local density and flux of matter and nongravitational energy-momentum, and $\chi^\mu=t^\mu+\OH\varphi^\mu$ generates the horizon, with $\varphi^\mu$ denoting the axial Killing vector. At quadratic order, gravitational perturbations contribute a non-negative horizon flux proportional to $\int_{\mathcal H}\sigma_{\mu\nu}\sigma^{\mu\nu}dV_{\mathcal H}$ \cite{GaoWald2001}, with the precise normalization depending on the horizon parameterization and canonical-energy convention. $\sigma_{\mu\nu}$ and $dV_{\mathcal H}$ are the shear tensor of the null generators of the horizon and the integration measure over a three-dimensional segment of the horizon, respectively. For a nonextremal black hole, $\kappa>0$ and, under the null energy condition and the usual regularity assumptions, these terms are non-negative \cite{HollandsWald2013}. Therefore,
\begin{equation}
 \mathcal D_H\geq 0
 \quad\Longrightarrow\quad\dot A\geq 0,\qquad\dot M_{\rm irr}\geq 0\label{eq:area-mirr-growth}
\end{equation}
with equality for a reversible transformation. At fixed flux, Eqs.~\eqref{eq:bzrates} and \eqref{eq:mirrtheorem} yield
\begin{equation}
 \mathcal D_H=\PBZ,\qquad \dot M_{\rm irr}=\frac{\PBZ}{4\kappa M_{\rm irr}}\label{eq:fixedflux}
\end{equation}
The rotational free-energy term decreases at twice the electromagnetic power, $-\OH\dot J_H=2\PBZ$. One half is carried outward from the horizon and one half is deposited irreversibly:
\begin{equation}  2\PBZ=\underbrace{\PBZ}_{\text{outward electromagnetic power}} +\underbrace{\frac{\kappa\dot A}{8\pi}}_{\text{horizon heat}}\label{eq:partition}
\end{equation}
For variable flux, while retaining the rates \eqref{eq:bzrates},
\begin{equation}
\dot M_{\rm irr}=\frac{1}{4\kappa M_{\rm irr}} \left(\PBZ-\frac{\phi_H}{2\pi}\dot\Phi\right)\label{eq:variableflux}
\end{equation}
Hence equality of horizon heating and outward electromagnetic power is a fixed-flux statement. In an accreting system, changes of $\Phi$ must be accompanied by the energy and angular momentum delivered by the disk. The apparent divergence as $\kappa\to0$ is not an infinite production of irreducible mass. The derivation assumes a nonextremal quasistationary sequence.  As the relaxation time $\kappa^{-1}$ diverges, this sequence ceases to be uniform unless $\mathcal D_H$ approaches zero sufficiently rapidly. The extremal limit requires a dynamical near-horizon analysis.

\noindent \emph{Integrated irreducible-mass growth}

\noindent The fixed-flux probe trajectory of Ref.~\cite{Wang2026} gives, from an initially extremal Kerr hole to a nonrotating endpoint,
\begin{equation}
 \frac{M_{H,f}}{M_{H,0}}=\frac{e^{1/4}}{\sqrt2}
 \label{eq:massratio}
\end{equation}
with $M_{H,0}$ denoting the initial horizon mass-energy, measured before the Blandford--Znajek extraction process begins. Since $A_0=8\pi M_{H,0}^2$ and $A_f=16\pi M_{H,f}^2$,
\begin{equation}
 \frac{A_f}{A_0}=\sqrt e,\qquad
 \frac{M_{\rm irr}^{(f)}}{M_{\rm irr}^{(0)}}=e^{1/4}=1.28403\ldots 
 \label{eq:integrated}
\end{equation}
Thus $M_H$ decreases by $9.21\%$, while $M_{\rm irr}$ increases from $M_{H,0}/\sqrt2$ to $0.90794M_{H,0}$, a change $\Delta M_{\rm irr}=0.20083M_{H,0}$. For the maximum-power trajectory $p=P/M_{H,0}=\sqrt{2/3}$ reported in Ref.~\cite{Wang2026}, $A_f/A_0\simeq1.567$ implies
$M_{\rm irr}^{(f)}/M_{\rm irr}^{(0)}\simeq1.252$.

\noindent \emph{Generalized Christodoulou decomposition}

\noindent For an isolated dyonic Kerr-Newman (KN) black hole carrying genuine global electric and magnetic Gauss charges $Q$ and $P$, electromagnetic duality gives the exact Christodoulou--Ruffini relation \cite{Christodoulou,Clement}
\begin{equation}
 M_{\rm KN}^2= \left(M_{\rm irr} +\frac{Q^2+P^2}{4M_{\rm irr}}\right)^2+\frac{J_{\rm KN}^2}{4M_{\rm irr}^2}.\label{eq:dyonicCR}
\end{equation}
This electrovac solution is not a split monopole: its magnetic charge is measured on a closed surface and is nonzero at infinity. In isolated-horizon mechanics, a Kerr--Newman-normalized mass functional may instead be assigned under the relevant Hamiltonian and boundary assumptions. The horizon angular momentum $J_H$ then depends not only on the intrinsic two-metric but also on the horizon rotation one-form and electromagnetic surface data \cite{Ashtekar2001}.

Wang's configuration is different. Its magnetic field reverses across the equatorial current sheet, and its net magnetic Gauss charge vanishes:
\begin{equation}\label{eq:zero-net-flux}
\frac{1}{4\pi}\int_{S^2}F = +P-P=0
\end{equation}
Thus $P=\Phi/(2\pi)$ is a sheet-supported hemispheric flux parameter, not the magnetic charge appearing in Eq.~\eqref{eq:dyonicCR}. We use
\begin{equation}
 M_H^2\equiv \left(M_{\rm irr}+\frac{P^2}{4M_{\rm irr}}\right)^2+\frac{J_H^2}{4M_{\rm irr}^2}\label{eq:CRflux}
\end{equation}
only as a conditional Kerr--Newman-normalized ansatz for Wang's horizon. Its validity requires a Hamiltonian derivation showing that the sheet-supported $P$ is an admissible horizon thermodynamic variable and that the equatorial boundary term is correctly included. It does not follow merely because $A$, $J_H$, and $\Phi$ can be assigned to a horizon cross section. Under this assumption, differentiating Eq.~\eqref{eq:CRflux} gives
\begin{equation}
 dM_H=4\kappa M_{\rm irr}\,d M_{\rm irr}+\OH dJ_H+\phi_H dP\label{eq:CRdiff}
\end{equation}
where $\OH=(\partial M_H/\partial J_H)_{M_{\rm irr},P}$ and $\phi_H=(\partial M_H/\partial P)_{M_{\rm irr},J_H}$. Using $dA=32\pi M_{\rm irr} dM_{\rm irr}$ and $dP=d\Phi/(2\pi)$ recovers Eq.~\eqref{eq:firstlaw}.  At fixed $P$ and $M_{\rm irr}$, reversible removal of all angular momentum gives
\begin{equation}
 M_{H,\min}=M_{\rm irr}+\frac{P^2}{4M_{\rm irr}}
\end{equation}
and therefore
\begin{equation}
E_{\rm rot}^{\max}=M_H-M_{\rm irr}-\frac{\Phi^2}{16\pi^2M_{\rm irr}}\label{eq:extractable}
\end{equation}
A BZ trajectory is irreversible because $M_{\rm irr}$ grows, and consequently delivers less energy than this reversible bound evaluated on the initial data.

\noindent\emph{Extracted energy.---}
The reversible bound and the actual BZ output are distinct. Consider an initially extremal magnetic Kerr--Newman-like state and define $p=P/M_{H,0}$. The extremal initial data used below are understood as the limiting values of a family of nonextremal quasistationary trajectories; the extremal limit is taken only after the spin-down equations have been integrated. Extremality then gives
\begin{equation}
a_0^2+P^2=M_{H,0}^2, \qquad M_{\rm irr}^{(0)} =\frac{M_{H,0}}{2}\sqrt{2-p^2}. \label{eq:initialmirr}
\end{equation}
Substitution in Eq.~\eqref{eq:extractable} gives the reversible efficiency
\begin{equation}
\eta_{\rm rev}(p)\equiv\frac{E_{\rm rot}^{\max}}{M_{H,0}c^2}=1-\frac{1}{\sqrt{2-p^2}}\label{eq:rev-efficiency}
\end{equation}
For extremal Kerr, $p=0$ and $\eta_{\rm rev}=1-1/\sqrt2=0.292893$.  For the leading maximum-power allocation $p^2=2/3$, Eq.~\eqref{eq:rev-efficiency} instead gives $\eta_{\rm rev}=1-\sqrt3/2=0.133975$. The energy extracted from the horizon along the dissipative BZ trajectory follows from
$\dot M_H=-\PBZ$:
\begin{equation}
 E_{\rm extr}^{H}=\int\PBZ\,dt=(M_{H,0}-M_{H,f})c^2\label{eq:horizonenergy}
\end{equation}
Eliminating time between the rates in Eq.~\eqref{eq:bzrates}, with $J_H=aM_H$, gives
\begin{equation}
\frac{dM_H}{da}=\frac{M_Ha}{2r_+^2+a^2},\qquad r_+=M_H+\sqrt{M_H^2-a^2-P^2}\label{eq:trajectory}
\end{equation}
The initial data are $M_H(a_0)=M_{H,0}$ and $a_0=M_{H,0}\sqrt{1-p^2}$; extraction ends at $a_f=0$. In the probe limit $p\to0$, Eq.~\eqref{eq:trajectory} integrates analytically to
$M_{H,f}/M_{H,0}=e^{1/4}/\sqrt2$.  Hence
\begin{equation}
\eta_H=\frac{E_{\rm extr}^{H}}{M_{H,0}c^2}=1-\frac{e^{1/4}}{\sqrt2}=0.0920569\label{eq:kerr-output}
\end{equation}
Therefore, the energy extracted from the horizon delivers $31.43\%$ of the reversible Kerr reservoir. The remaining difference is encoded in the increase of irreducible mass:
\begin{equation}
\underbrace{0.292893M_{H,0}}_{M_{H,0}-M_{\rm irr}^{(0)}} =\underbrace{0.0920569M_{H,0}}_{E_{\rm extr}^{H}/c^2} +\underbrace{0.200836M_{H,0}}_{M_{\rm irr}^{(f)}-M_{\rm irr}^{(0)}} \label{eq:kerrpartition}
\end{equation}
This integrated identity should not be confused with the instantaneous equal-power relation \eqref{eq:partition}; the conversion between horizon heat and $d M_{\rm irr}$ is weighted by the evolving $\kappa$ and $M_{\rm irr}$. For finite fixed $P$, numerical integration of Eq.~\eqref{eq:trajectory}
gives the efficiencies in Table~\ref{tab:energies}.  At the leading maximum-power allocation $p=\sqrt{2/3}$,
\begin{equation}
\frac{M_{H,f}}{M_{H,0}}=0.953316,\qquad \eta_H=\frac{E_{\rm extr}^{H}}{M_{H,0}c^2}=0.0466844\label{eq:maxpower-output}
\end{equation}
Thus, $4.668\%$ of the initial horizon mass is extracted, corresponding to $34.85\%$ of the reversible rotational reservoir for the same initial horizon data.

\begin{table}[b]
\caption{Reversible rotational-energy bound and energy removed from the horizon for initially extremal states. The BZ values follow from Eq.~\eqref{eq:trajectory} at fixed
$P$.}
\label{tab:energies}
\begin{ruledtabular}
\begin{tabular}{c c c c}
$p=P/M_{H,0}$ & $\eta_{\rm rev}$ & $\eta_H$
& $E_{\rm extr}^{H}/E_{\rm rot}^{\max}$\\
\hline
$\to0$ & $0.29289$ & $0.09206$ & $0.3143$\\
$1/\sqrt3$ & $0.22540$ & $0.07334$ & $0.3254$\\
$2/3$ & $0.19822$ & $0.06561$ & $0.3310$\\
$\sqrt{2/3}$ & $0.13397$ & $0.04668$ & $0.3485$\\
\end{tabular}
\end{ruledtabular}
\end{table}
Restoring units, the probe and maximum-power-flux outputs are, for one solar mass $M_\odot$, respectively\footnote{The probe limit is $p=P/M_{H,0}\to0$: $P_{\rm BZ}\to0$ and the spin-down time diverges, but the integrated extracted fraction remains finite. The maximum-power case, $p=\sqrt{2/3}$, maximizes the leading instantaneous power rather than the lifetime-integrated energy.}

\begin{align}\label{eq:physicalunits}
&E_{\rm extr}^{H,\,\mathrm{probe}}=1.65\times10^{53} \left(\frac{M_{H,0}}{M_\odot}\right){\rm erg},\nonumber\\
&E_{\rm extr}^{H,\,\mathrm{max}}=8.35\times10^{52} \left(\frac{M_{H,0}}{M_\odot}\right){\rm erg}
\end{align}
The situation is summarized in Table~\ref{tab:probe_max_power}. Although the instantaneous power vanishes in the probe limit, the spin-down timescale diverges, leaving a finite lifetime-integrated output of $9.206\%$ of $M_{H,0}c^2$. By contrast, the initial maximum-power configuration produces a larger instantaneous luminosity but contains a smaller rotational reservoir, yielding $4.668\%$ of $M_{H,0}c^2$ over its lifetime.
\begin{table}[t]
\caption{\label{tab:probe_max_power}
Comparison of the probe limit with the initial configuration that maximizes the leading-order instantaneous BZ power. Here $p\equiv P/M_{H,0}$ and $\eta_H\equiv E_{\rm extr}^{H}/(M_{H,0}c^2)$. The maximum-power configuration does not maximize the lifetime-integrated extracted energy.}
\begin{ruledtabular}
\begin{tabular}{lcccc}
Regime
& \(p\)
& \(a_0/M_{H,0}\)
& \(P_{\rm BZ}\)
& \(\eta_H\) \\
\hline
Probe
& \(\to 0\)
& \(\to 1\)
& \(\to 0\)
& \(0.09206\) \\
Maximum power
& \(\sqrt{2/3}\)
& \(1/\sqrt{3}\)
& \(P_{\rm BZ}^{\max}\)
& \(0.04668\) \\
\end{tabular}
\end{ruledtabular}
\end{table}

\noindent \textit{Quasilocal versus global mass}.
We have deliberately reserved  $E_{\rm jet}^{\infty} \equiv \int P_\infty\,du $ for the energy actually received at null infinity. The quantities are related by the exterior energy balance,
\begin{equation}\label{en}
E_{\rm extr}^{H} = E_{\rm jet}^{\infty} +\Delta E_{\rm EM}^{\rm ext} +E_{\rm diss}^{\rm ext} +E_{\rm boundary}
\end{equation}
so that $E_{\rm jet}^{\infty}=E_{\rm extr}^{H}$ only when no net energy remains stored or dissipated in the exterior, and no additional boundary contribution is present.

\noindent For a true Kerr--Newman magnetic charge, $P$ is a Gauss charge defined on a closed two-surface and measurable at infinity. In the split monopole, the hemispheric fluxes have opposite signs, and the total flux through the full sphere vanishes [Eq.~\eqref{eq:zero-net-flux}]. The quantity $\Phi_N=\int_{S_H^N}F=2\pi P$ is a nonzero hemispheric flux, not a net magnetic charge measured at infinity.
The nonzero hemispheric flux is supported by an equatorial current sheet. We reserve $M_{\rm ADM}$ and $J_{\rm ADM}$ for charges at spatial infinity and $M_{\rm Bondi}$ for the mass at null infinity; none of these is identified with $M_H$ or $J_H$ without an additional global argument. A global completion has schematically, in geometrized units,
\begin{equation}
 M_{\rm ADM}=M_H+E_{\rm sheet}+E_{\rm disk}+E^{\rm ext}_{EM}+E_{\rm int}\label{eq:globalmass}
\end{equation}
where $M_{\rm ADM}$ is the total mass-energy of the complete spacetime measured at spatial infinity. The terms $E_{\rm sheet}$ and $E_{\rm disk}$ denote the energies associated with the equatorial current sheet and its physical disk completion, $E_{\rm EM}^{\rm ext}$ is the electromagnetic energy stored outside the horizon, and $E_{\rm int}$ collects interaction, equatorial boundary, and possible renormalization contributions. The ideal static junction of Ref.~\cite{Wang2026} has vanishing surface stress-energy, but this does not prove that every term in Eq.~\eqref{eq:globalmass} vanishes for the rotating, current-carrying, asymptotically matched engine. Consequently, replacing $M_H$ and $J_H$ by $M_{\rm ADM}$ and $J_{\rm ADM}$ is an additional global statement. Establishing it requires a Hamiltonian or covariant-phase-space derivation including the current sheet, disk boundary, and exterior wind. The same issue applies to treating $\Phi$ as a globally conserved thermodynamic charge in an extremality budget.

\noindent \emph{Discussion.}
\noindent Wang derived the secular evolution of a self-gravitating BZ engine, but left its horizon thermodynamics undetermined. We have supplied this missing information. At fixed flux and impedance matching, the rotational work divides instantaneously into equal outward electromagnetic power and irreversible horizon heating. The integrated evolution is not equally divided: in the weak-field limit, the horizon area grows by the exact factor $\sqrt e$, while the energy removed from the horizon is $9.206\%$ of its initial mass energy. For the leading maximum-power spin--flux allocation, the extracted fraction is $4.668\%$. These are the principal new results of this work.

The analysis also identifies precisely what is local and what remains global. Equation~\eqref{eq:mirrtheorem} is a quasilocal horizon balance under the assumed isolated-horizon first law, and Eq.~\eqref{eq:CRflux} is a conditional Kerr--Newman-normalized functional for the sheet-supported flux. Because Wang's extracting solution is a slow-rotation, quasistationary expansion with nonstandard near-zone asymptotics, neither $M_H=M_{\rm ADM}$ nor $E_{\rm extr}^{H}=E_{\rm jet}^{\infty}$ follows automatically. Establishing those equalities requires the current-sheet boundary term, a controlled matching across the light cylinder, and the Bondi flux at null infinity.

This separation makes Wang's idealized engine a useful laboratory for the energetics of newly formed rotating black holes. In particular, the explicit distinction between extractable horizon energy and energy received by an outflow provides a controlled starting point for confronting black-hole energy budgets with high-energy transients, including the GeV emission discussed for GRB~220101A and GRB~090510 \cite{ruffini2}. The present calculation establishes the horizon contribution; its astrophysical completion is the global flux problem identified above.

\noindent\textbf{Acknowlwdgents}

\noindent We thank the Minister of Foreign Affairs and International Cooperation of Italy (MAECI) for financial support.

\noindent \textbf{Author Contributions}

\noindent R.R. and G.S. contributed equally to this work.

\noindent\textbf{Data Availability Statement}

\noindent The data supporting the findings of this study are available from the authors upon reasonable request.

%\begin{acknowledgments}
%We thank [names] for useful discussions.  This work was supported by [funding information].
%\end{acknowledgments}


\begin{thebibliography}{99}

\bibitem{Christodoulou} D. Christodoulou, Phys. Rev. Lett. \textbf{25}, 1596 (1970).

\bibitem{ruffini1} D. Christodoulou and R. Ruffini, Phys. Rev. D \textbf{4}, 3552 (1971).

\bibitem{Hawking1971} S. W. Hawking, Phys. Rev. Lett. \textbf{26}, 1344 (1971).

\bibitem{ruffini} R. Ruffini and J.R. Wilson,  Phys. Rev. D, \textbf{12}(10), 2959-2962 (1975).

\bibitem{BZ1977} R. D. Blandford and R. L. Znajek, Mon. Not. R. Astron. Soc. \textbf{179}, 433 (1977).

\bibitem{MT1982} D. Macdonald and K. S. Thorne, Mon. Not. R. Astron. Soc. \textbf{198}, 345 (1982).

\bibitem{Wang2026} Y. Wang, \emph{An Exact Engine for Black-Hole Jets}, submitted to Astronomy Commmunications (2026).

\bibitem{GaoWald2001} S. Gao and R. M. Wald, Phys. Rev. D \textbf{64}, 084020 (2001).

\bibitem{HollandsWald2013} S. Hollands and R. M. Wald, Commun. Math. Phys. \textbf{321}, 629 (2013).

\bibitem{Ashtekar2001} A. Ashtekar, C. Beetle, and J. Lewandowski, Phys. Rev. D \textbf{64}, 044016 (2001).

\bibitem{Clement} G. Cl\'ement and D. Gal'tsov, Phys. Lett. B \textbf{773}, 290 (2017).

\bibitem{ruffini2} R. Ruffini \textit{et al.}, \textit{GRB 220101A: a most energetic $10^{54}$ erg long GRB triggered by two supernovae 3.5 seconds apart}, submitted to Astrophysical Journal (APJ) (2026).

\end{thebibliography}
\end{document}